\documentclass[traditabstract, rnote]{aa}
\usepackage{natbib}
\bibpunct{(}{)}{;}{a}{}{ , }
\usepackage[dvipdfm]{graphicx}
\usepackage{amsmath}
\usepackage{booktabs}
\usepackage{txfonts}

\begin{document}
\title{A coordinate-independent technique for detecting globally inhomogeneous flat topologies}
\date{Received / Accepted}
\author{H. Fujii}
\keywords{cosmology: theory - cosmology: large scale structure of Universe}
\authorrunning{H. Fujii}
\institute{Institute of Astronomy, School of Science, University of Tokyo, 2-21-1, Osawa, Mitaka, Tokyo 181-0015, Japan}
\titlerunning{A coordinate-independent technique for detecting globally inhomogeneous flat  topologies}
\abstract{
A flat Universe model supported by recent observations has 18 possible choices for its overall topology. To detect or exclude these possibilities is one of the most important tasks in modern cosmology, but it has been very difficult for globally inhomogeneous ones because of a long-time calculation. In this brief paper we provide an object-based 3D method to overcome the problem, as an extension of  Fujii \& Yoshii (2011a). 
Though the test depends on the observer's location in the universe, this method drastically reduces calculation times to constrain inhomogeneous topologies, and will be useful in exhaustively constraining the size of the Universe.
}

\maketitle

\section{Introduction}

Theory of modern cosmology is based on Einstein's General Relativity, and recent observations favor a $\Lambda$-CDM cosmology with vanishing curvature (e.g.,  $\Omega_{\mathrm{tot}}=1.0050^{+0.0060}_{-0.0061}$ from \textit{WMAP}+BAO+SN data, by Hinshaw et al. 2009), which successfully describes the observed properties such as the cosmic structure formation, the cosmic microwave background (CMB) anisotropies, and the accelerating expansion. 
However, General Relativity does not distinguish two spaces with the same curvature but with different topologies. Though the curvature of our Universe is exactly zero, we still have 18 choices for the overall topology of the Universe (Nowacki 1934), e.g., the multiconnected 3-torus $\mathbb{T}^3$ with finite volume.

A multiconnected space is a quotient space of the simply connected space with the same curvature ($\mathbb{S}^3, \mathbb{E}^3, \ \mathrm{or} \ \mathbb{H}^3$), by a holonomy group $\Gamma$. An observer in the space would see it as a finite or infinite $2K$-polyhedron (the \textit{Dirichlet domain}) whose $K$ pairs of faces are glued by the holonomies. There exist the second, third, and more, geodesics between a given object and the observer, which across the Dirichlet domain. As a result, multiple images of single objects, often referred to as ``ghosts", can be observed in a multiconnected space (for details, see, e.g.,  Lachi\`eze-Rey and Luminet 1995).

Based on this prediction, many object-based works for constraining cosmic topology were carried out, i.e., they searched for periodic and symmetrical patterns made by ghosts of galaxies, galaxy clusters,  or active galactic nuclei (e.g., Fagundes \& Wichoski 1987; Demia\'nski \& \L apucha 1987; Lehoucq et al. 1996; Roukema 1996; Uzan et al. 1999; Weatherley et al. 2003; Marecki et al. 2005; Menzies \& Mathews 2005). However, their methods are valid only for particular topologies and/or, not sensitive enough to constrain spaces that are comparable to the observed region in size. The method by Roukema (1996) has both the generality and the high sensitivity, but the algorithm is not so sophisticated and the calculations take very long times. 

One of the recent trends is to use the \begin{em}circles-in-the-sky\end{em} method (Cornish et al. 1998) that is to search for intersections of the last-scattering surface and the faces of our Dirichlet domain. They appear as circles with the same temperature fluctuation pattern in the CMB map since they are physically identical. This method can detect any topologies, but checking all the possibilities requires an extremely long-time calculation because of the many free parameters: radius of the matched circles and celestial positions of centers of the two circles. Such exhaustive analyzations have not carried out yet, and various authors have searched for matched circles limiting to antipodal or nearly antipodal ones, using the \begin{em}WMAP\end{em} satellite's data. However, they obtained diverse results (e.g., Aurich 2008; Roukema et al. 2008; Cornish et al. 2004; Key et al. 2007; Bielewicz \& Banday 2011),  which suggests that there are methodological problems. 


 One way to break down this situation is to use observational data that is independent of the CMB observations, e.g., astronomical objects such as galaxies and quasars. This motivates us to revisit the object-based 3D methods, and we developed a method that can constrain any of the 18 flat topologies with simple algorithms, and is much more sensitive to topological signatures than the preceding ones (Fujii \& Yoshii 2011a, hereafter FY11a). Nevertheless, it also has a similar problem in checking all the possible topologies: calculation time becomes very long. The aim of this brief paper is to provide a technique to overcome  this problem. Throughout the paper we consider flat universes.
 
\section{Method}

\subsection{Mathematical background and definitions}
Mathematical classification of the holonomy groups for flat spaces was completed by Nowacki (1934). Any holonomy $\gamma$ can be written as $\gamma = \gamma_{T} \gamma_{NT}$, where $\gamma _{T}$ is a parallel translation and $\gamma_{NT}$ is one of an identity, $n$-th turn rotations (for $n=2, 3, 4,$ and $6$), and a reflection. 
Those spaces whose holonomy groups include only translations ($\gamma_{NT} = id$) are called (globally) homogeneous, otherwise called (globally) inhomogeneous. We investigate the latter topologies in this paper.

We borrow a convenient way of writing the holonomies that is to use a 4D coordinate system $(w,x,y,z)$ where the simply connected 3-Euclidean space $\mathbb{E}^3$ is represented as a hyperplane $w=1$, so that a usual 3D vector $(x,y,z)$ is represented as a 4D vector $\vec x = (1,x,y,z)$. Every holonomy $\gamma$  is also written as a $4\times 4$ matrix, e.g., a half-turn corkscrew motion is written as: 
\[ \gamma = U \left( \begin{tabular}{cccc}1 & 0 & 0 & 0 \\ 0 & -1 & 0 & 0 \\ 0 & 0 & -1 & 0 \\ 0 & 0 & 0 & 1  \end{tabular} \right )U^{-1} + \left( \begin{tabular}{cccc}1 & 0 & 0 & 0 \\ $L_1$ & 0 & 0 & 0 \\ $L_2$ & 0 & 0 & 0 \\ $L_3$ & 0 & 0 & 0  \end{tabular} \right ),\]
where $U$ is a $4 \times 4$ matrix representing the choices of the coordinate systems, which reduces to $U=id$ if we choose our $z$-axis to be parallel to the rotational axis (hereafter, we call this straight line the \textit{fundamental axis} of the holonomy, denoted by $l_{fun}$). The unit vector  $\vec n_{fun}$ that is parallel to $l_{fun}$ is called the \textit{fundamental vector}; we identify two antipodal vectors $\vec n_{fun}$ and $-\vec n_{fun}$. For a holonomy of reflection, fundamental axis $l_{fun}$ is parallel to the reflectional plane. 

In the rest of this paper, unless otherwise stated we continue to consider the same half-turn corkscrew motion as an example.


\subsection{Summary of the 3D method of FY11a}
In this section we review the FY11a's object-based 3D method for detecting cosmic topology (see the paper for details). Our assumption is that the Universe has the spatial section with zero curvature, as suggested by recent observations. This assumption is also favored by theory of the quantum creation of the universe (Linde 2004).

If a pair of comoving objects $\vec x_1$ and $\vec x_1'$ are linked by a holonomy $\gamma$, we have by definition
\begin{equation}
\vec x_1' = \gamma \vec x_1 = \gamma_{T} \gamma_{NT} \vec x_1 = \gamma_{NT} \vec x_1 + \vec L, 
\end{equation}
where $\vec L=(1,L_1,L_2,L_3)$ is the translational vector. 
Detecting such topological twins requires the parameter searching for 5 parameters: the translational vector $\vec L$ and the fundamental vector $\vec n_{fun}$. To eliminate $\vec L$ we search for two pairs of ghosts (called a \textit{topological quadruplet}) $[(\vec x_1, \vec x_2), (\vec x_1', \vec x_2 '$)] such that
\begin{equation}
\vec x_1' = \gamma \vec x_1 = \gamma_{T} \gamma_{NT} \vec x_1 = \gamma_{NT} \vec x_1 + \vec L, 
\end{equation}
\begin{equation}
\vec x_2' = \gamma \vec x_2 = \gamma_{T} \gamma_{NT} \vec x_2 = \gamma_{NT} \vec x_2 + \vec L.
\end{equation}
Such a quadruplet always satisfies
\begin{equation}
\vec x_2' - \vec x_1' = \gamma_{NT} (\vec x_2 - \vec x_1),
\end{equation}
independent of $\vec L$. If our $z$-axis is parallel to the fundamental axis $l_{fun}$ (hence $U = id$), the following relations hold:
\begin{equation}
x_2 ' - x_1 ' = -(x_2 - x_1), \label{con1} \end{equation}\begin{equation} 
y_2 ' - y_1 ' = -(y_2 - y_1), \label{con2} \end{equation}\begin{equation}
z_2 ' - z_1 ' = z_2 - z_1, \label{con3} \end{equation}
since $\gamma$ is a half-turn corkscrew motion here.

\begin{figure}[!htb]
\centering
{\includegraphics[width=8cm,trim=0 0 0 0]{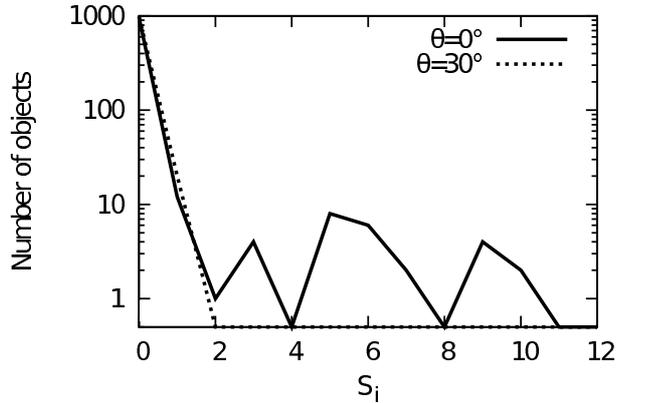}}
\caption{Solid line: the result using the $z$-axis parallel to $\vec n_{fun}$. Broken line: the result using the $z$-axis that is at the angle of $30^{\circ}$ with $\vec n_{fun}$. The vertical scale is linear from 0 to 1 and logarithmic from 1 to 1000. The observer stands at the distance of $3$ Gpc from $l_{fun}$ (see section 3.2).} 
\label{fig1} 
\end{figure}

The FY11a's scheme is to search for quadruplets $[(\vec x_i, \vec x_j),(\vec x_k, \vec x_l)]$ that simultaneously satisfy the following three conditions:
\begin{itemize}
\item[1. ] separation condition:  $|\vec x_i - \vec x_j| = |\vec x_k - \vec x_l|$.
\item[] This condition is common to all holonomies, as holonomies are isometries that preserve distance. Preceding works such as Roukema (1996) and Uzan et al. (1999) also used this mathematical property of holonomies.
\item[]
\item[2. ] vectorial condition:  eq. (5)-(7).
\item[] This is for a half-turn corkscrew motion; for other types of holonomies, see FY11a. Similar condition for a translation was used in Marecki et al. (2005).
\item[]
\item[3. ] lifetime condition: $|t_i - t_k|, |t_j - t_l| < t_{\mathrm{life}}$.
\item[ ] The variables $t_i, t_j, \cdots$ are cosmic times of objects $\vec x_i, \vec x_j, \cdots$, respectively, and $t_{\mathrm{life}}$ is the typical lifetime of objects. This condition is important when considering short-lived objects, e.g., active galactic nuclei.
\end{itemize}
Due to these multi-filters, only $\lesssim 10$ pairs of ghosts among the total $>1000$ objects are sufficient for this test.
However, eq. (\ref{con1})-(\ref{con3}) hold only for the specific coordinate system, so we still have to perform the parameter searching for 2 parameters, the fundamental vector $\vec n_{fun}$. To our dismay, this requires over about $10^6$ trials that takes a very long time. These strong and weak points of the FY11a's method is seen in Fig \ref{fig1}, where $s_i$ is the number of quadruplets that include the object $\vec x_i$ as its members and satisfy all the conditions 1-3 (see FY11a for details). When using the $z$-axis parallel to $\vec n_{fun}$, there can be seen some bumps in the histograms that are constituted by topological ghosts, while such bumps are not seen when the $z$-axis deviates from $\vec n_{fun}$. The property of the simulated catalog used here is described in section 3.1.

\subsection{Fundamental-vector-searching method}

\begin{figure}[!htb]
\centering
{\includegraphics[width=5cm, trim=0 0 0 0]{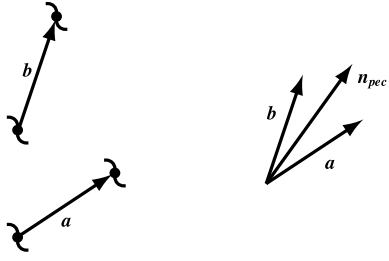}}
\caption{Even a nontopological quadruplet seems to be linked by a half-turn corkscrew motion when we choose the $z$-axis parallel to its peculiar axis $\vec n_{pec} \propto \vec a + \vec b$.} 
\label{fig2} 
\end{figure}

One may want to have a coordinate-independent filter to extract topological quadruplets, but it is found to be impossible. 
Consider an arbitrary quadruplet $[(\vec x_i, \vec x_j),(\vec x_k, \vec x_l)] $ that already satisfies the separation condition and the lifetime condition. It then also satisfies the vectorial condition eq. (\ref{con1})-(\ref{con3}) if we choose the $z$-axis parallel to the vector $ \vec a + \vec b$, where $\vec a = \vec x_j - \vec x_i$ and $\vec b= \vec x_l - \vec x_k$ (Fig \ref{fig2}). The unit vector $\vec n_{pec} \propto \vec a + \vec b$ is called the \textit{peculiar vector} of the quadruplet, for a half-turn corkscrew motion. 

Therefore it is impossible to judge whether a given quadruplet is topological, however, if there really exists the fundamental axis (and vector) in the Universe, the number of quadruplets whose peculiar vectors are parallel to it should be larger than stochastically expected. This is the essential idea of the coordinate-independent technique described here. Detailed procedure, hereafter we call the ``\textit{fundamental-vector-searching} (FVS)" method,  is as follows.
\begin{itemize}
\item[1. ] The celestial sphere is divided into $12N_{\mathrm{side}}^2$ equal area pixels. Since we identify  two antipodal vectors, the number of pixels used is $6N_{\mathrm{side}}^2$.
\item[2. ] The quadruplets that satisfy both the separation condition and the lifetime condition are selected.
\item[3. ] For each selected quadruplet, its peculiar vector $\vec n_{pec}$ and the celestial pixel containing $\vec n_{pec}$ within it is calculated.
\item[4. ] For each object $\vec x_i$, the pixels containing more than $s_{\mathrm{min}}$ peculiar axes of the quadruplets that include $\vec x_i$ are flagged. 
\item[5. ] Each pixel is assigned an integer, the number of times of being flagged in the previous step.
\end{itemize}
These steps practically correspond to counting the number of objects with $s_{i} > s_{\mathrm{min}}$ for all the possible $z$-axes. Note that the choices of the parameters $N_{\mathrm{side}}$ and $s_{\mathrm{min}}$ are not independent; a small $N_{\mathrm{side}}$ corresponds to large pixels and large stochastic noises, so $s_{\mathrm{min}}$ should be chosen to be large enough. 
HEALPix scheme (G\'orski et al. 2005) is used in pixelization of the celestial sphere.

We consider the half-turn corkscrew motion as an example throughout this paper, but the treatments on the other types of holonomies are similar except for the step 3.
The peculiar axis of the quadruplet $[(\vec x_i, \vec x_j),(\vec x_k, \vec x_l)] $ for each type of holonomies can be calculated as follows.

\

\noindent \textit{1. \ $n$-th turn corkscrew motions}

In this case the fundamental axis is defined to be parallel to the rotational axis. The apparent angle between $\vec a$ and $\vec b$, therefore, should be $2 \pi / n$ when seen from its peculiar vector $\vec n_{pec}$. Such a vector is obtained by rotating the unit vector parallel to $\vec a+\vec b$ around the axis parallel to $\vec a- \vec b$ by the angle of 
\begin{equation}
\alpha = \pm \arcsin \Biggl ( \frac{\tan \phi}{\tan \frac{\pi}{n}} \Biggr ),
\end{equation}
where $\phi$ is half the intrinsic angle between $\vec a$ and $\vec b$. The quadruplet with  $\phi > \pi /n$ has no peculiar axes, which never occurs when $n=2$ (half-turn corkscrew motion).

\

\noindent \textit{2. \ glide reflections}

In this case the fundamental axis is defined to be perpendicular to the reflectional plane, hence the peculiar vector is the unit vector parallel to $\vec a - \vec b$.

\section{Simulations and Discussions}

\subsection{Details of simulated catalog}

We generated a full-sky catalog of toy quasars to test the FVS method. The properties of the catalog is given here.
\begin{itemize}
\item The standard $\Lambda$-CDM cosmology ($\Omega_\mathrm{m}=0.27, \Omega_{\Lambda}=0.73, H_0=71$ km/sec/Mpc) is used, so that the local geometry is Euclidean. 
 \item A half-turn space topology is assumed, such that the detectable holonomies are only two half-turn corkscrew motions. The translational distance of these detectable holonomies is $L=10$ Gpc.
 \item the comoving number density of quasars obeys the empirical law for the luminous quasars (Fan et al. 2001b).
 \item Quasar luminosity evolution is simplified such that they emit radiation with constant luminosity during the fixed duration $t_{\mathrm{life}}=10^8$ yr.
 \item The peculiar motion of quasars is simplified to move with constant speed $v = 500$ km/sec and with randomly chosen directions.
 \item Other natures of quasars (such as clustering, activity cycle, and anisotropic morphology) and the technical uncertainties are all ignored.
 \end{itemize}
 We acknowledge that these simple assumptions are not allowed for practical applications. Simulations in a more realistic situation will be a subject of the coming paper.

\subsection{Main results: detection of the fundamental axis}

First we applied the FY11a's method to a toy quasar catalog seen from an observer standing at a comoving distance of $3$ Gpc from $l_{fun}$. The catalog contains 1010 objects at redshifts $z>4.7$, including 12 pairs of ghosts. Two choices of coordinate systems, $\theta=0^{\circ}$ and $30^{\circ}$ where $\theta$ is the angle between the chosen $z$-axis and $\vec n_{fun}$, were used. These results are given in Fig \ref{fig1} and already discussed; topological signal appears only for the former case ($\theta=0^{\circ}$), and we have to change the $z$-axis as free parameters to detect globally inhomogeneous topologies (see also section 5.2.1 of FY11a). This takes an extremely long time, and it is our motivation for this work.

Next we applied the FVS method to the same catalog using $N_{\mathrm{side}}=500$ and $s_{\mathrm{min}}=2$. If there is a topological quadruplet, its peculiar vector $\vec n_{pec}$ ideally coincides with the fundamental vector $\vec n_{fun}$, but not in practice due to peculiar motions. The resolution parameter $N_{\mathrm{side}}$ is determined so as to cover this positional deviations. The cutoff parameter $s_{\mathrm{min}}=2$ is a reasonable choice because the quadruplets with $s_i>2$ seem to rarely occur by chance as can be seen in Fig \ref{fig1}. 

\begin{figure}[!htb]
\centering
{\includegraphics[clip,width=4.45cm,trim=0 0 0 0]{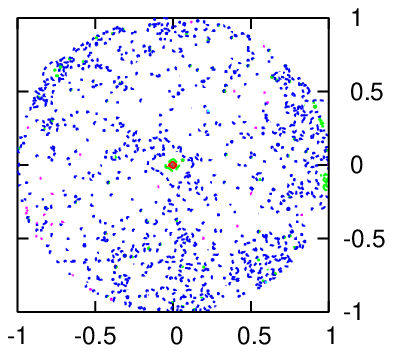}}
{\includegraphics[clip,width=4.45cm,trim=0 0 0 0]{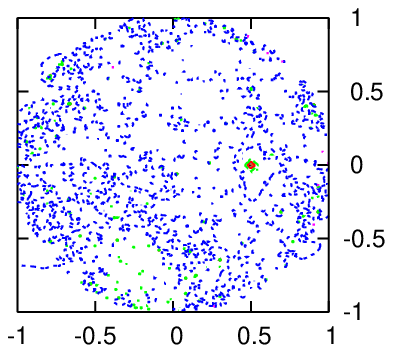}}
\caption{The fundamental vector $\vec n_{fun}$ appears as a sharp peak in both cases; left: $\theta=0^{\circ}$, right: $\theta=30^{\circ}$. The observer stands at the distance of 3 Gpc from the fundamental axis $l_{fun}$. Plots outside the unit disk are fake made by square-shaped meshing of the region $[-1,1] \times [-1,1]$.} 
\label{fig3} 
\end{figure}

The results were smoothed and contoured in 2D maps (Fig \ref{fig3}), where the half celestial sphere is orthogonally projected on the $xy$-plane. As expected, a pixel containing the fundamental axis has larger counts than the other pixels whose counts are purely stochastic, independently of the coordinate systems. 

This single calculation takes about a few ten minutes with a present-day ordinary personal computer, and is equivalent to performing about $\sim10^6$ trials of the FY11a's method, taking about one year. This is the notable advantage of the FVS method over the other methods, e.g., the CMB-based circles-in-the-sky method that cannot constrain all the possible topologies in a reasonable time.

\subsection{Limits of the method: observer's location}


The results of the previous section are based on the somewhat idealized assumptions. In practice, there are various effects to be considered correctly: physical properties of astronomical objects and observational uncertainties. These effects will be treated in the coming papers and not here, however, there is another important factor that affects this test; the limit of the FVS method depends on the observer's location in the universe. 

To see this effect, let us consider a topological quadruplet $[(\vec x_i, \vec x_j),(\vec x_i ', \vec x_j ')]$, which satisfies
\begin{equation}
\vec a_{\perp} = - \vec b_{\perp} + \vec \varepsilon_{\perp},
\end{equation}\begin{equation}
\vec a_{\parallel} = \vec b_{\parallel} - \vec \varepsilon_{\parallel},
\end{equation}
where $\vec a = \vec x_j - \vec x_i$, $\vec b = \vec x_j ' - \vec x_i '$, and $\vec \varepsilon$ represents the deviation due to peculiar motions. The lower indices $\perp$ and $\parallel$ represent the vector components perpendicular and parallel to $ l_{fun}$, respectively. The peculiar vector of this quadruplet is 
\begin{equation}
\vec n_{pec} \propto \vec a + \vec b \propto \vec n_{fun} + \frac{\vec \varepsilon_{\parallel} + \vec \varepsilon_{\perp}}{2|\vec a_{\parallel}|}. \label{devi}
\end{equation}
If the observer stands near the fundamental axis $l_{fun}$, the faces of his Dirichlet domain are nearly perpendicular to $l_{fun}$.
On the other hand, ghosts of the short-lived objects such as quasars and starburst galaxies, which are luminous and suitable for high-redshift observations, appear near the faces of the Dirichlet domain. As a result, $\vec a_{\parallel}$ becomes small and slight changes in position due to peculiar velocity induce a large deviation of $\vec n_{pec}$ from $\vec n_{fun}$.

\begin{figure}[!htb]
\centering
{\includegraphics[clip,width=5.1cm,trim=0 0 0 0]{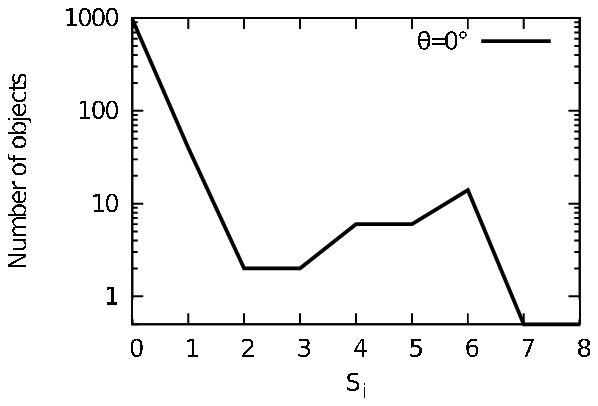}}
{\includegraphics[clip,width=3.7cm,trim=0 0 0 0]{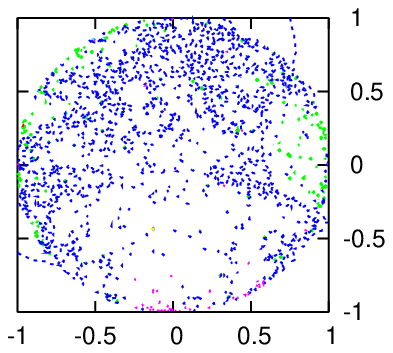}}
\caption{In the case where the observer stands along $l_{fun}$, the FVS method does not work well (right), and a tremendous parameter searching of the FY11a's method is necessary (left). } 
\label{fig4} 
\end{figure}

Therefore, if the Milky Way is situated near $l_{fun}$, the FVS method can no longer detect the topological signatures and a troublesome parameter searching is needed. For example, we generated a toy quasar catalog seen from the observer standing along $l_{fun}$. The catalog contains 1040 objects including 14 pairs of ghosts. In Fig \ref{fig4}, it can be seen that the FVS method does not detect the fundamental vector, while the FY11a's method successfully detects topological signature though in practice a long-time calculation to search for $\vec n_{fun}$ is needed.

Now we discuss the limitation of the FVS method a little more in detail. In order that there be a sharp peak, both the following conditions must be satisfied:
\begin{itemize}
\item[1. ] Sufficient number of topological ghosts are in the catalog.
\item[2. ] All the peculiar vectors $\vec n_{pec}$ of the topological quadruplets are lying in one (or a few) pixel.
\end{itemize}
To satisfy the condition 2, the pixel size should be larger than the typical deviation from $\vec n_{fun}$ of $\vec n_{pec}$'s of topological quadruplets, which can be estimated from eq. (\ref{devi}). Then the typical number of $\vec n_{pec}$'s of nontopological quadruplets (including a given object $\vec x_i$)  contained in one pixel can be calculated. The condition 1 claims that this stochastic noise must be sufficiently smaller than the number of topological ghost pairs.

\begin{figure}[!htb]
\centering
{\includegraphics[clip,width=8.1cm,trim=0 0 0 0]{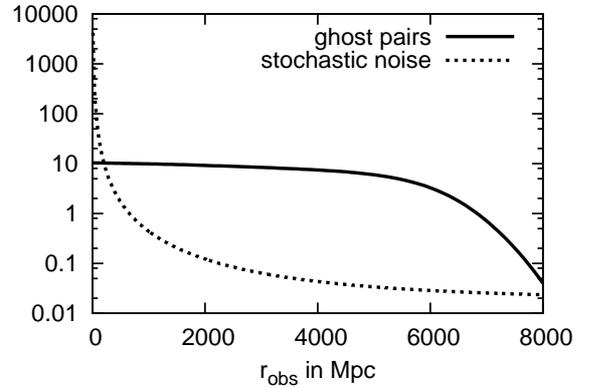}}
\caption{The FVS method is not valid if stochastic noise $>$ ghost pairs. The universe's typical size is assumed to be $L=10$ Gpc here.} 
\label{fig5} 
\end{figure}

Fig \ref{fig5} shows the rough estimation of these quantities as a function of $r_{\mathrm{obs}}$, the observer's distance from $l_{fun}$, in the situation described in section 3.1. The typical size of the universe is fixed to $L=10$ Gpc for simplicity, though it should be treated as a variable. The FVS method is valid as long as the stochastic noise is small enough compared to the ghost pairs, i.e., $1 \ \mathrm{Gpc} \lesssim r_{\mathrm{obs}} \lesssim 6 \ \mathrm{Gpc}$, which is consistent with the previous results ($r_{\mathrm{obs}}=3 \ \mathrm{Gpc}$ and $0$ Gpc). When there is no sharp peaks as in Fig \ref{fig4}, we can conclude that $r_{\mathrm{obs}} \lesssim 1$ Gpc or $r_{\mathrm{obs}} \gtrsim 6$ Gpc for $L=10$ Gpc, and the constraints on $r_{\mathrm{obs}}$ for other values of $L$ also can be obtained similarly.

\subsection{Prospects}

We introduced the FVS method that, along with the FY11a's method, is useful for constraining topology of the universe with zero curvature. Especially, the former method will provide the first observational constraints on the inhomogeneous topologies, which take an unrealistically long time when using other CMB-based or object-based methods.

However, for practical application we need more realistic simulations. Then our next work will focus on these issues: detailed treatments on physical properties of astronomical objects and technical uncertainties of observations. When this accomplished, we will be able to apply the FY11a's method and the FVS method to the observed catalogs of our real Universe. Recent and future large-scale survey projects, such as 2dFGRS (Croom et al. 2004), SDSS (Schneider et al. 2010), Large Synoptic Survey Telescope (LSST) and the Joint Astrophysics Nascent Universe Satellite (JANUS), provide sufficient data for this test. Formidable progress in techniques for measuring photometric redshifts suggests that the spectroscopic surveys are not necessarily needed.



\acknowledgement{I gratefully want to thank Y. Yoshii for his various supports, useful discussions and constructive suggestions. I also thank M. Doi, T. Minezaki, T. Tsujimoto, T. Yamagata, T. Kakehata, K. Hattori, and T. Shimizu for useful discussions and suggestions. Some of the results in this paper were derived using the HEALPix package.}

\nocite{*}
\bibliography{FVS_print}

\end{document}